\documentclass[submitting]{nst}

\usepackage{subfigure,dcolumn}
\usepackage[T2A,T1]{fontenc}
\usepackage[russian,english]{babel}
\usepackage{epstopdf}
\usepackage{mhchem}
\usepackage{upgreek}
\usepackage{amsmath}

\usepackage{listings}
\begin{document}

\title{Multi-frequency Point Supported LLRF Front-end for CiADS Wide-bandwidth Application}

\author{Qi Chen}
\affiliation{Institute of Modern Physics, Chinese Academy of Sciences, Lanzhou 730000, China}
\affiliation{School of Nuclear Sciences and Technology, University of Chinese Academy of Sciences, Beijing 100049, China}

\author{Zheng Gao}
\affiliation{Institute of Modern Physics, Chinese Academy of Sciences, Lanzhou 730000, China}

\author{Zheng-Long Zhu}
\affiliation{Institute of Modern Physics, Chinese Academy of Sciences, Lanzhou 730000, China}

\author{Zong-Heng Xue}
\affiliation{Institute of Modern Physics, Chinese Academy of Sciences, Lanzhou 730000, China}

\author{Yuan He}
\email[Corresponding author, ]{hey@impcas.ac.cn}
\affiliation{Institute of Modern Physics, Chinese Academy of Sciences, Lanzhou 730000, China}

\author{Xian-Wu Wang}
\affiliation{Institute of Modern Physics, Chinese Academy of Sciences, Lanzhou 730000, China}

\begin{abstract}

The China initiative Accelerator Driven System, CiADS, physics design adopts $162.5 \,\mathrm{MHz}$, $325 \,\mathrm{MHz}$, and $650 \,\mathrm{MHz}$ cavities, which are driven by the corresponding radio frequency (RF) power system, requiring frequency translation front-end for the RF station. For that application, a general-purpose design front-end prototype has been developed to evaluate the multi-frequency point supported design feasibility. The difficult parts to achieve the requirements of the general-purpose design are reasonable device selection and balanced design. With a carefully selected low-noise wide-band RF mixer and amplifier to balance the performance of multi-frequency supported down-conversion, specially designed local oscillator (LO) distribution net to increase isolation between adjacent channels, and external band-pass filter to realize expected up-conversion frequencies, high maintenance and modular front-end general-purpose design has been implemented. Results of standard parameters show an $R^2$ value of at least 99.991\% in the range of $-60 \sim 10\,\mathrm{dBm}$ for linearity, up to $18\,\mathrm{dBm}$ for P1dB, and up to $89\,\mathrm{dBc}$ for crosstalk between adjacent channels. The phase noise spectrum is lower than $80\,\mathrm{dBc}$ in the range of $0 \sim 1\,\mathrm{MHz}$, and cumulative phase noise is $0.006^\circ$; amplitude and phase stability are $0.022\%$ and $0.034^\circ$, respectively.

\end{abstract}


\maketitle

\section{Introduction}\label{sec.I}

In recent years, particle accelerators have found applications in many fields. They are used in nuclear and particle physics research, in industrial applications such as ion implantation and lithography, in biological and medical research with synchrotron light sources, and in material science and medical research with spallation neutron sources; they have are also used in radiotherapy, food sterilization, waste treatment, etc.\cite{leeaccelerator}. In modern large particle accelerators, several hundred meters to kilometers long linacs or storage rings are adopted to attain higher energy or larger beam currents~\cite{cern00,nlc00,imp00}. Large quantities of different $\beta$ resonance cavities working at different frequencies were employed to accelerate heavy ions to gain the desired beam power. The development of Accelerator driven sub-critical system (ADS) facilities like ABC-ADDT-ATW-AFCI (USA), JAERI-ADS (Japan 2004), HYPER (Korea), INR (Russia), etc.\cite{gulevich2008}, and the CiADS facility is a similar application that contains a high-intensity proton super-conducting linacs~\cite{ciads00,liu2017physics}.

According to the CiADS physics design, $162.5 \,\mathrm{MHz}$ is used at the front end of the ion accelerator to allow larger longitudinal acceptance at low energies, and also to facilitate component machining, and at intermediate energies, the frequency is increased  to $325 \,\mathrm{MHz}$ to benefit from a higher accelerating gradient while minimizing the size of the cavities in this area. Similarly, $650 \,\mathrm{MHz}$ is used for later cavities, and is also called the frequency jump \cite{eshraqi2012beam}. The frequency jump structure is widely adopted for obtaining high intensity proton beams in the design of the ADS \cite{verma2019major,li2013physics}.

\begin{figure*}[!htb]
\includegraphics[width=.95\hsize]{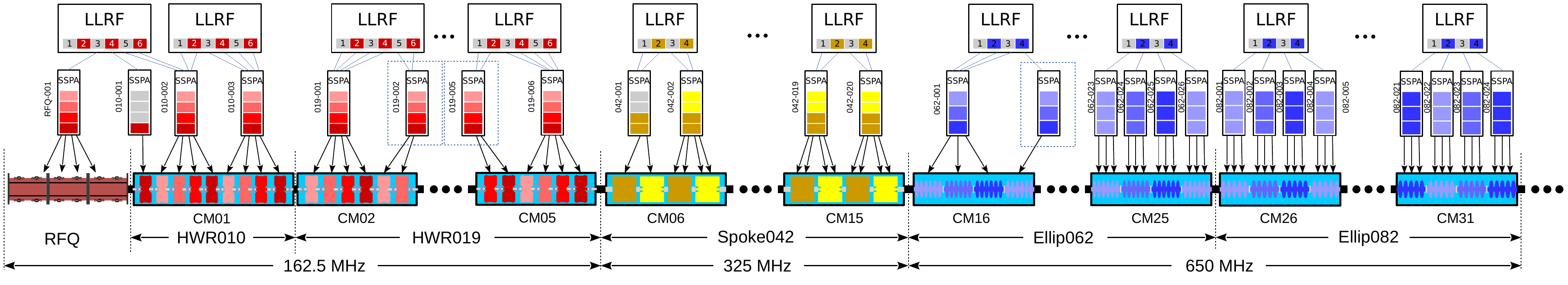}
\caption{(Color online) SSPA and LLRF distribution for CiADS}
\label{fig:SSPAandLLRFDistribution}
\end{figure*}

CiADS will work under $10\,\mathrm{mA}$ current continuous wave (CW) mode after several upgrade phases. CiADS has critical requirements on beam trip time, for current operation experience on $25\,\mathrm{MeV}$ CAFe (The China ADS Front End Superconducting Demo Linac), most beam trips were caused by the Radio Frequency (RF) system, similar to European Synchrotron Radiation Facility (ESRF) \cite{hardy2003accelerator}. To consider the requirement of flexibility to upgrade the RF power during the beam power upgrading phases, the structure of a single resonance cavity driven by an RF Solid State Power Amplifier (SSPA) controlled with a Low Level RF (LLRF) system was chosen (see Fig.~\ref{fig:SSPAandLLRFDistribution}). For that application method, a large amount of the LLRF system will be employed for CiADS, accordingly, the same amount of RF front-end is required to finish frequency translation. For the benefits of the digital algorithm, which can easily reach high quality and precision control procedure, digitizer and frequency conversion modules have been widely introduced in modern LLRF systems \cite{corredoura1999architecture,schnase2005control,simrock2005digital}. The main reason for frequency translation is satisfying the accuracy requirement and reasonable price for ADC (Analog to Digital Converter), and not directly sampling at very high frequency according to the Nyquist-Shannon sampling theorem. For the frequency jump application, frequency translation becomes a challenge for RF front-end design, especially for multi-frequency supported design. The Mean Time to Repair (MTTR) and maintenance backup cost will increase for customized single frequency front-end, obviously, if all expected frequencies can be supported will decrease unnecessary cost on maintenance and backup. For implementing general-purpose structure on the board level, a multi-frequency supported RF front-end prototype board was developed and tested to evaluate feasibility of this design.

\section{Basic parameters}\label{sec.II}

Frequency jump has already existed once in CAFe between Taper HCM6 and SCM6 according to the physical lattice design, from the $162.5\,\mathrm{MHz}$ jump to $325\,\mathrm{MHz}$ \cite{liu2017physics}. In the $500\,\mathrm{MeV}$ CiADS physical lattice design, except for working at the $162.5\,\mathrm{MHz}$ Half Wave Resonance (HWR) cavity, increased working at the $325\,\mathrm{MHz}$ Spoke cavity and $650\,\mathrm{MHz}$ Elliptical cavity will be employed to deliver higher beam power~\cite{he2019development,wang2019status}. Amplitude and phase stability RMS values are $0.1\%$ and $0.1^\circ$ respectively for the LLRF control system proposed according to physical design after beam dynamic simulation. Based on the requirements of accuracy, the error source was initially analyzed, including the correlation and uncorrelation error. Correlation error comes from detuning caused by microphonics and the Lawrence force, and phase drift originates from reference line variation by the ambient temperature; uncorrelation error is mainly introduced by equipment noise.

Since the clock jitter of the directly sampled ADC and DAC reduces the SNR of the receiving system~\cite{kurosawa2002sampling}, the method of RF translating to the intermediate frequency (IF) (also known as superheterodyne) is often used for higher frequencies. CiADS intends to down-convert RF to IF at 25 MHz (for digitization) or up-convert IF to RF. The number of channels needs to meet the requirements of at least three down-conversions (cavity pickup signals, power forward, and reflected coupling signals) for a single cavity.

The basic signal margin is determined based on the requirement of stability for amplitude and phase, as well as other signal requirements. Amplitude stability of $0.1\%$ corresponding an SNR value is $-60\,\mathrm{dBc}$, while phase stability of $0.1^\circ$ is $-71.13\,\mathrm{dBc}$. Precision measurement of the cavity pickup signal with proper attenuation calibration can calculate $E_{\mathrm{pk}}$, if required at $10\%$ accuracy for the measurement value, then the linearity coefficient of determination $R^2$ for the down-conversion channel should no less than $99\%$. More than $10\,\mathrm{dBm}$ are required to ensure the signal strength at the front-end RF input port, P1dB, which is no less than $10\,\mathrm{dBm}$ correspondingly. The IF signal range determined by the ADC and DAC input or output signals full-scale value, regularly $\pm 1\,\mathrm{V}$ differential pair, for $50\,\mathrm{\Omega}$ impedance is $10\,\mathrm{dBm}$, and for linearity and attenuation consideration, is set to $7\,\mathrm{dBm}$. The up-conversion channel output signal maximum $10\,\mathrm{dBm}$ is temporary.

\section{Design and implementation}

The implementation difficulties of general-purpose multi-frequency point supported design RF front-end may come from the wide-band devices selection, the noise figure control, the PCB layout and tracing, the crosstalk compression design.

\subsection{Principle of wide-band design}

To minimize the effects of wide-band supporting design, especially for filter quantity and position design, mixer selection should be considered carefully. To improve the quality of the RF signal, a filter will be located at the RF input port or LO input port normally, that will limit the RF bandwidth. For wide-band devices such as the mixer and amplifier selection, the performance is not perfect for all frequencies, which may affect the linearity and adjacent channel crosstalk performances. In addition, the PCB layout and tracing for the high frequency RF signal is more difficult than for low frequency signals, since the impedance matching characters are more sensitive for high RF signal crosstalk control.

Another consideration for multi-frequency point supported design is the filter. Only a low-pass filter (LPF) was designed, located after the amplifier ADA4937 to filter out frequency higher than IF for down-conversion channel, while for up-conversion channel, a same LPF was used before the mixer to filter out the bias frequency from the DAC. The external BPF was designed to filter out the mirror frequency for up-conversion channel. For better LO input purity, an LPF device position was also reserved for a fixed specific frequency if only a single frequency were needed.

\subsection{Device selection and SNR control}

For multi-frequency point supported design, the key devices include the mixer and amplifier, which should be wide-band supported and have balanced performances at expected frequencies. The mixer AD8342 and amplifier ADA4937 were chosen after carefully comparing the parameters of IP3, noise figure, and bandwidth with other similar devices. The AD8342~\cite{analog:ad8342} is a high performance, broadband active receive mixer from low frequency to $3.8 \,\mathrm{GHz}$. The ADA4937~\cite{analog:ada4937} is low noise, ultra-low distortion, high-speed differential amplifier from $-3 \,\mathrm{dB}$ at the bandwidth of $1.9 \,\mathrm{GHz}$. Both devices have excellent third-order intercept point (IP3) guaranteeing linear performance for the down-conversion channel. For the up-conversion channel, a single-ended amplifier from Mini-Circuits was selected for higher gain performance for the RF output level.

The calculation details on the noise figure with a single channel of SNR from $-20 \sim 10 \mathrm{dBm}$ is $60 \sim 90 \mathrm{dB}$ for the chosen devices, according to the Friis’ Formula and noise factor definition. The bandwidth chosen during the calculation of the noise figure is based on the overall loop bandwidth at the operating frequency, while the working bandwidth of the superconducting cavity is only hundreds of hertz, and the tuning range is only a few or several tens of kilohertz.

\subsection{Block diagram and PCB layout}

\begin{figure}[!htb]
  \includegraphics[width=.95\hsize]{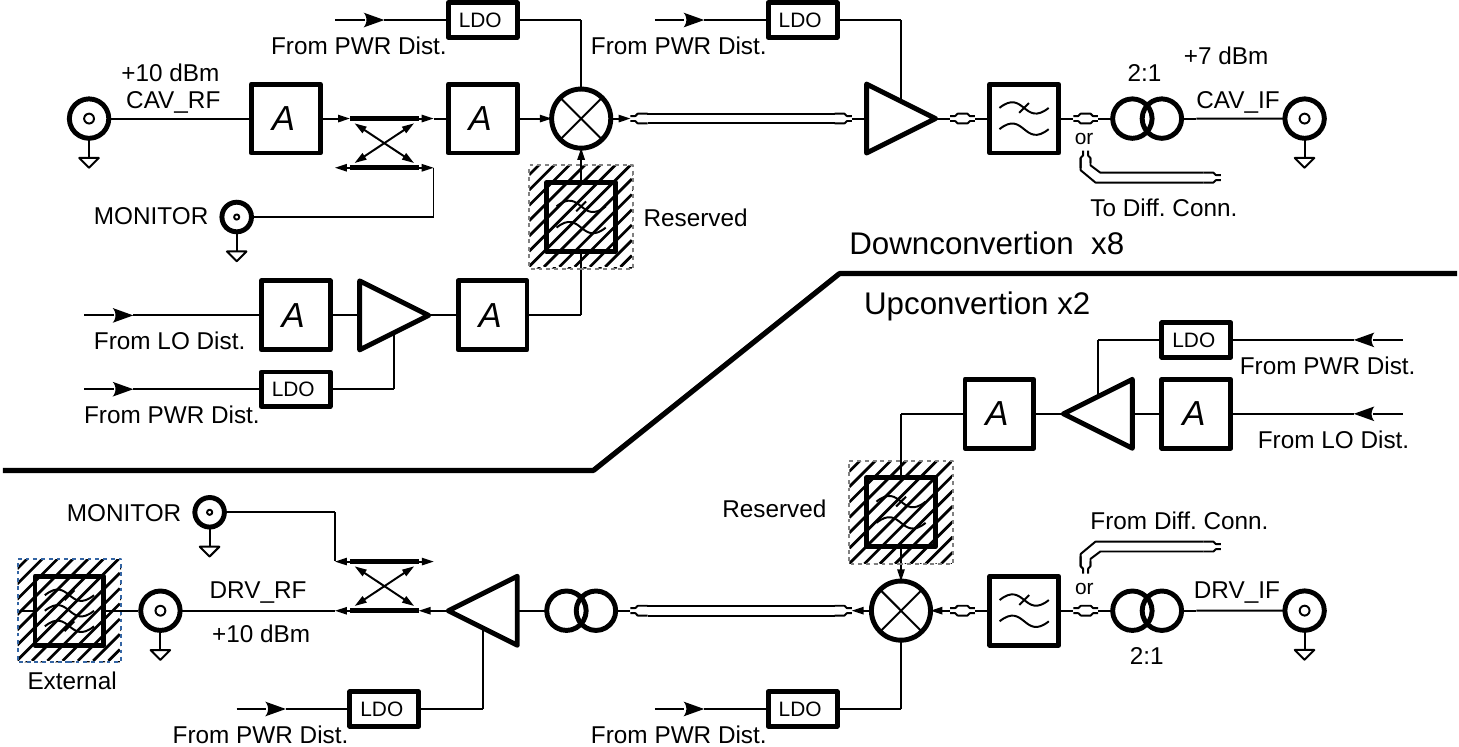}
\caption{(Color online) Schematic for up and down conversion channels}
\label{fig:IMPRFFE-Detailed}       
\end{figure}

\begin{figure}[!htb]
  \includegraphics[width=.95\hsize]{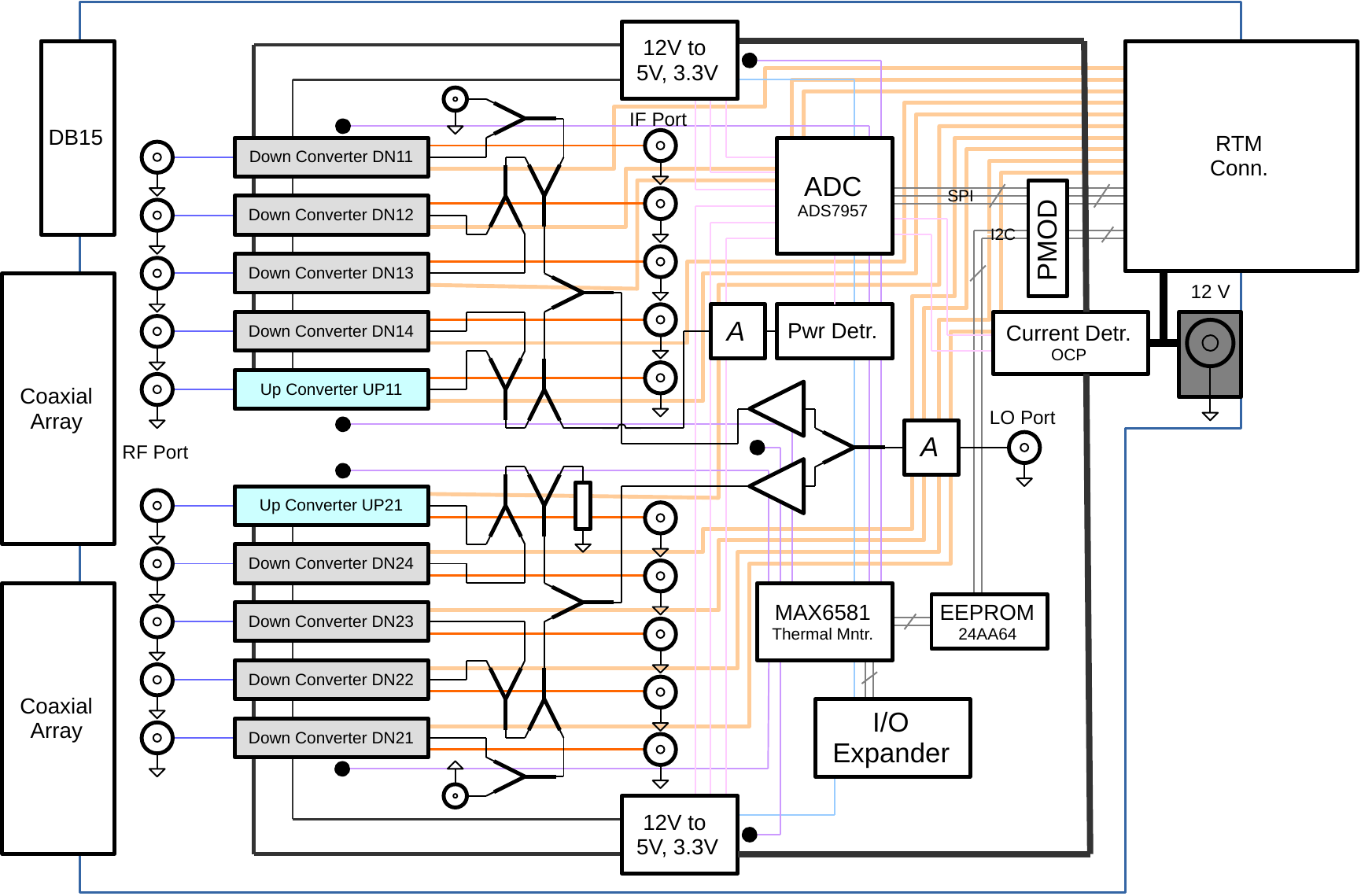}
\caption{(Color online) Block diagram of front-end}
\label{fig:LLRF_Front-end}       
\end{figure}

\begin{figure}[!htb]
  \includegraphics[width=.95\hsize]{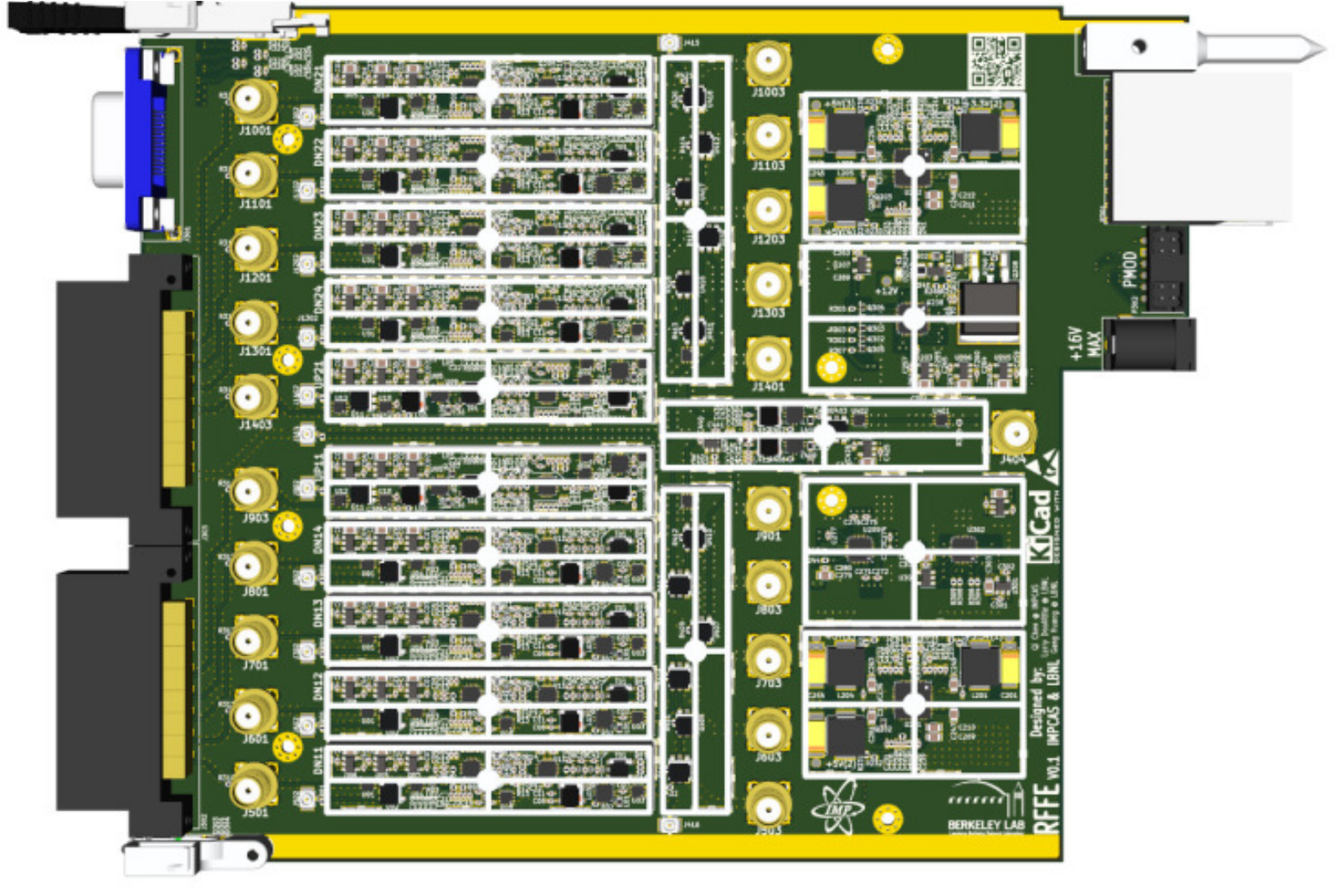}
\caption{(Color online) MTCA.4 RTM compatible design PCB layout of front-end}
\label{fig:IMPRFFE_PCB}       
\end{figure}

The front-end board is designed as two mirrored up-and-down frequency conversion units for better PCB layout and coplanar waveguide tracing to decrease the crosstalk (see Fig.~\ref{fig:LLRF_Front-end}). Each unit has four-way down-conversion channels and a one-way up conversion channel. It uses the IF port differential signal design (can also be converted to single-ended signals to use or test). The shielded differential connector is conveniently connected to the digitizer as shown in Fig.~\ref{fig:LLRF_Front-end}. The PCB design is compatible with the size and interface functions of the MTCA.4 RTM board. The corresponding interface and basic shape are shown in Fig.~\ref{fig:LLRF_Front-end} and the final PCB layout is shown in Fig.~\ref{fig:IMPRFFE_PCB}. For down-conversion channels, the maximum 10 dBm RF signal is passed through a directional coupler and attenuator with 0 dBm LO through a mixer to obtain a 25 MHz signal, followed by a low-pass filter to obtain a differential IF signal, which is connected to mTCA.4 RTM standard differential connector via two shorting resistors, or through a 2:1 transformer to a single-ended IF output via an SMA connector (as shown in Fig.~\ref{fig:IMPRFFE-Detailed}).

\begin{figure*}[!htb]
  \includegraphics[width=.95\hsize]{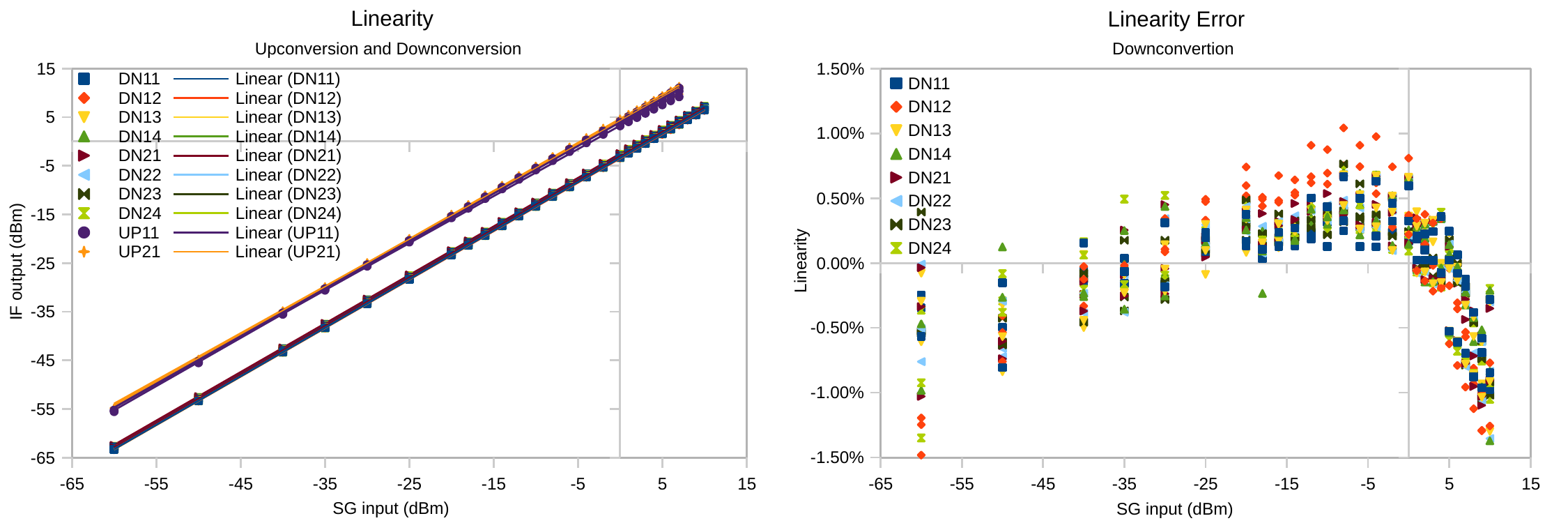}
\caption{(Color online) Linearity and error for each down-conversion channel at the expected frequency}
\label{fig:Linearity}       
\end{figure*}

\begin{figure}[!htb]
  \includegraphics[width=0.95\hsize]{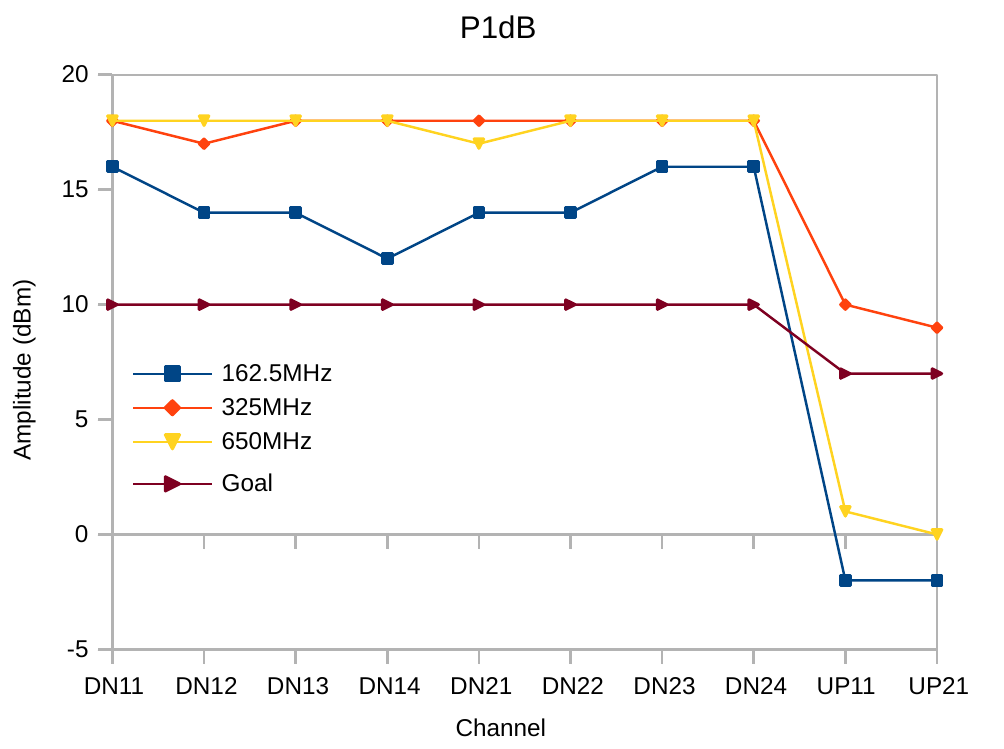}
\caption{(Color online) P1dB for each down-conversion channel at the expected frequency}
\label{fig:P1dB}       
\end{figure}

The multi-channel integrated design reduces the size of the LLRF system, reduces group delay, and simplifies internal connections, but greatly increases the difficulty of the PCB layout and RF signal isolation design. If the designs of the ADC, DAC, and RF front-end are integrated, the tracing around the ADC will increase, which will reduce the signal integrity and increase the complexity of the PCB layout. Therefore, the digital and RF front-end separation design is adopted~\cite{mavric2008design}. Based on the calculation of the noise figure for the selected devices, the operating points of the mixer and amplifier are determined, thereby ensuring the feasibility of the system multi-channel integrated general-purpose design.

Other design aspects include the generation and distribution for power supplies, the acquisition and transmission for monitoring signals. The power supply is converted from $12 \,\mathrm{V}$ DC to $6 \,\mathrm{V}$ and $4.3 \,\mathrm{V}$ DC through two power buck converters, and then converted to the required $5 \,\mathrm{V}$ and $3.3 \,\mathrm{V}$ DC power supply for each main chip with LDOs to minimize the buck converter output ripple (as shown in Fig.~\ref{fig:IMPRFFE-Detailed}). Considering that the necessary signals need to be monitored, a multi-channel slow ADC is employed for the voltage and current of the power supply, and LO signal power level. The temperature-monitoring module is designed to monitor the up and down conversion channels. The monitoring signals, and the extension of some of the state control signals are transmitted over the I2C or SPI bus.

\section{Test results and discussion}

\begin{figure}[!htb]
  \includegraphics[width=.95\hsize]{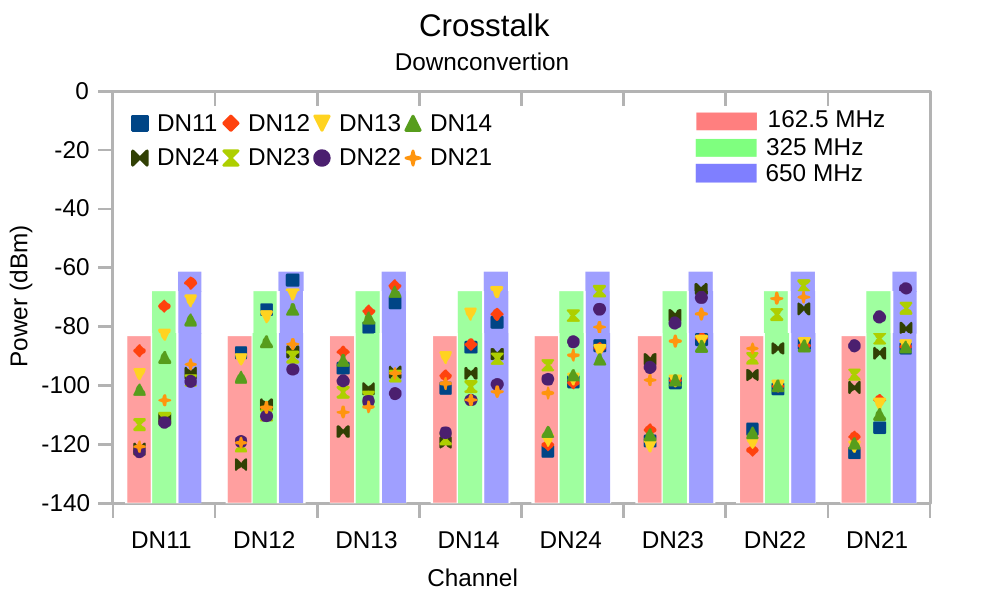}
\caption{(Color online) Crosstalk at expected frequency}
\label{fig:Crosstalk}       
\end{figure}

\begin{figure}[!htb]
  \includegraphics[width=.95\hsize]{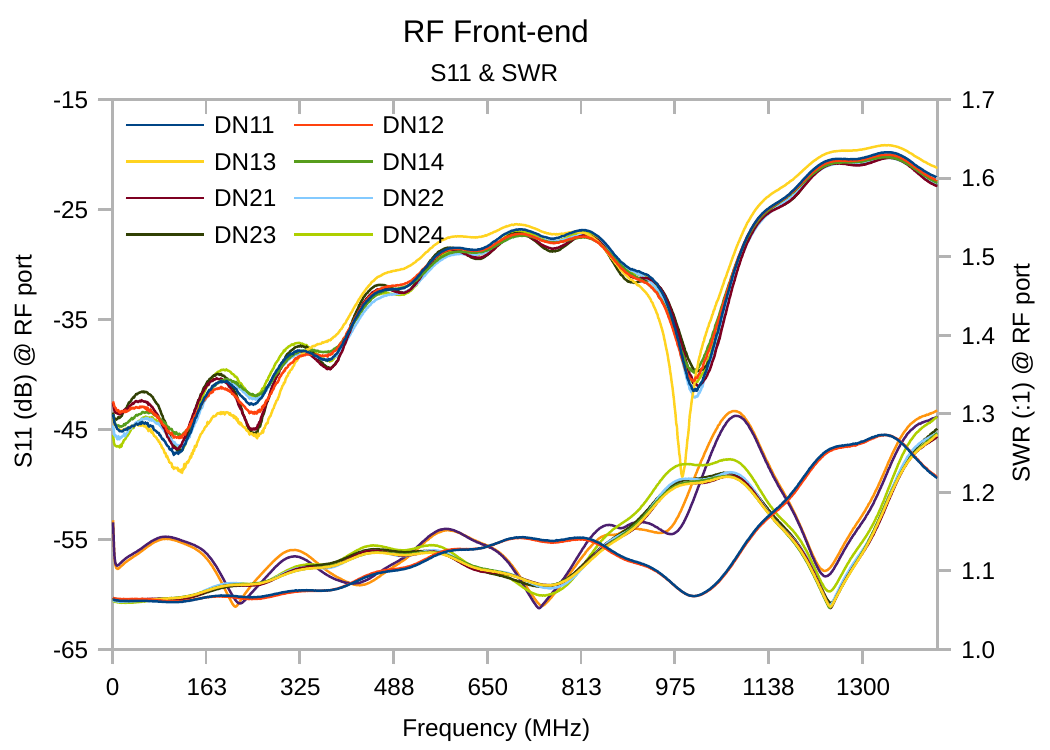}
\caption{(Color online) S11 and SWR per channel}
\label{fig:S11SWR1}       
\end{figure}

\begin{figure*}[!htb]
  \includegraphics[width=.95\hsize]{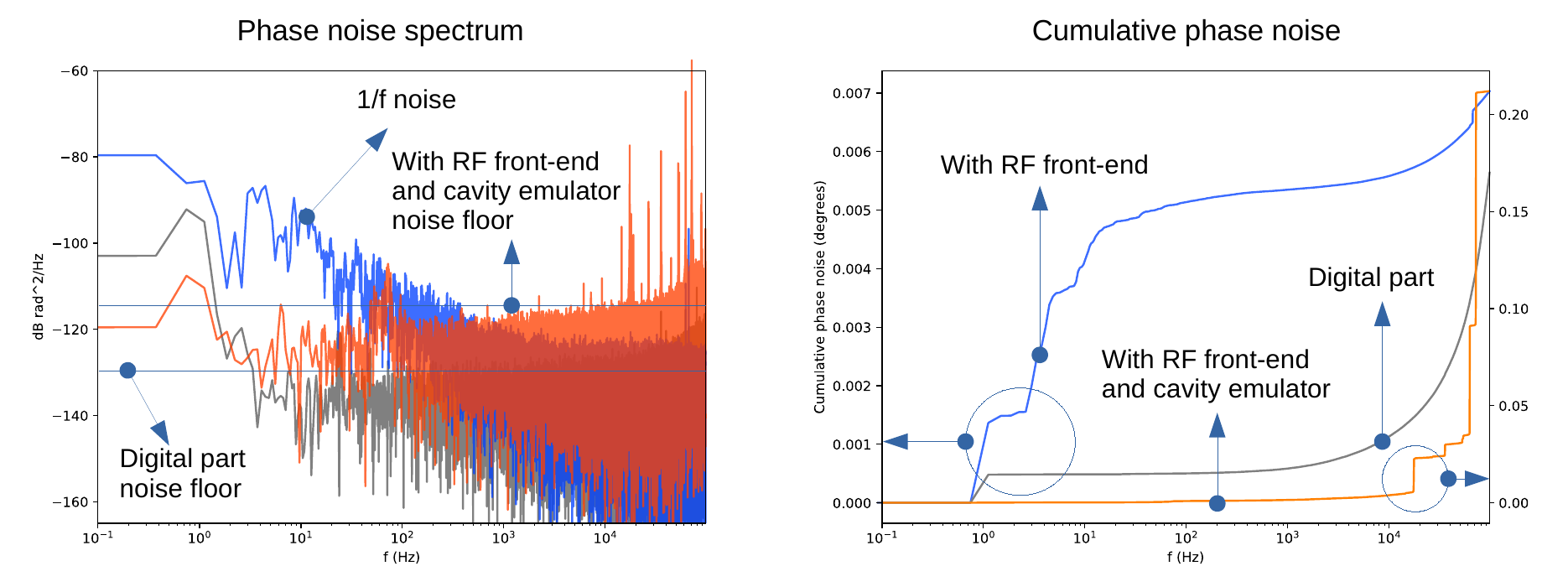}
\caption{(Color online) Phase noise test results for different loops}
\label{fig:Phasenoise}       
\end{figure*}

\begin{figure*}[!htb]
  \includegraphics[width=.95\hsize]{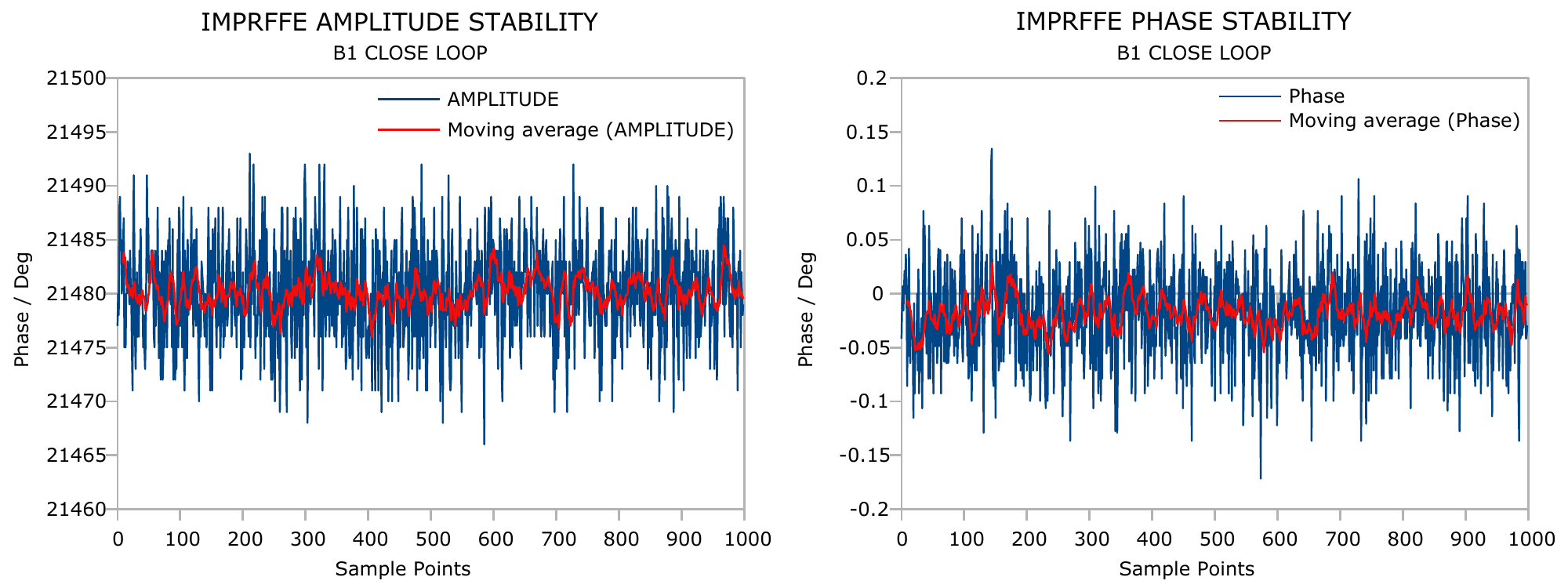}
\caption{(Color online) Stability test result of buncher cavity LLRF close-loop with front-end}
\label{fig:B1Closeloop}       
\end{figure*}

\begin{figure}[!htb]
  \includegraphics[width=.95\hsize]{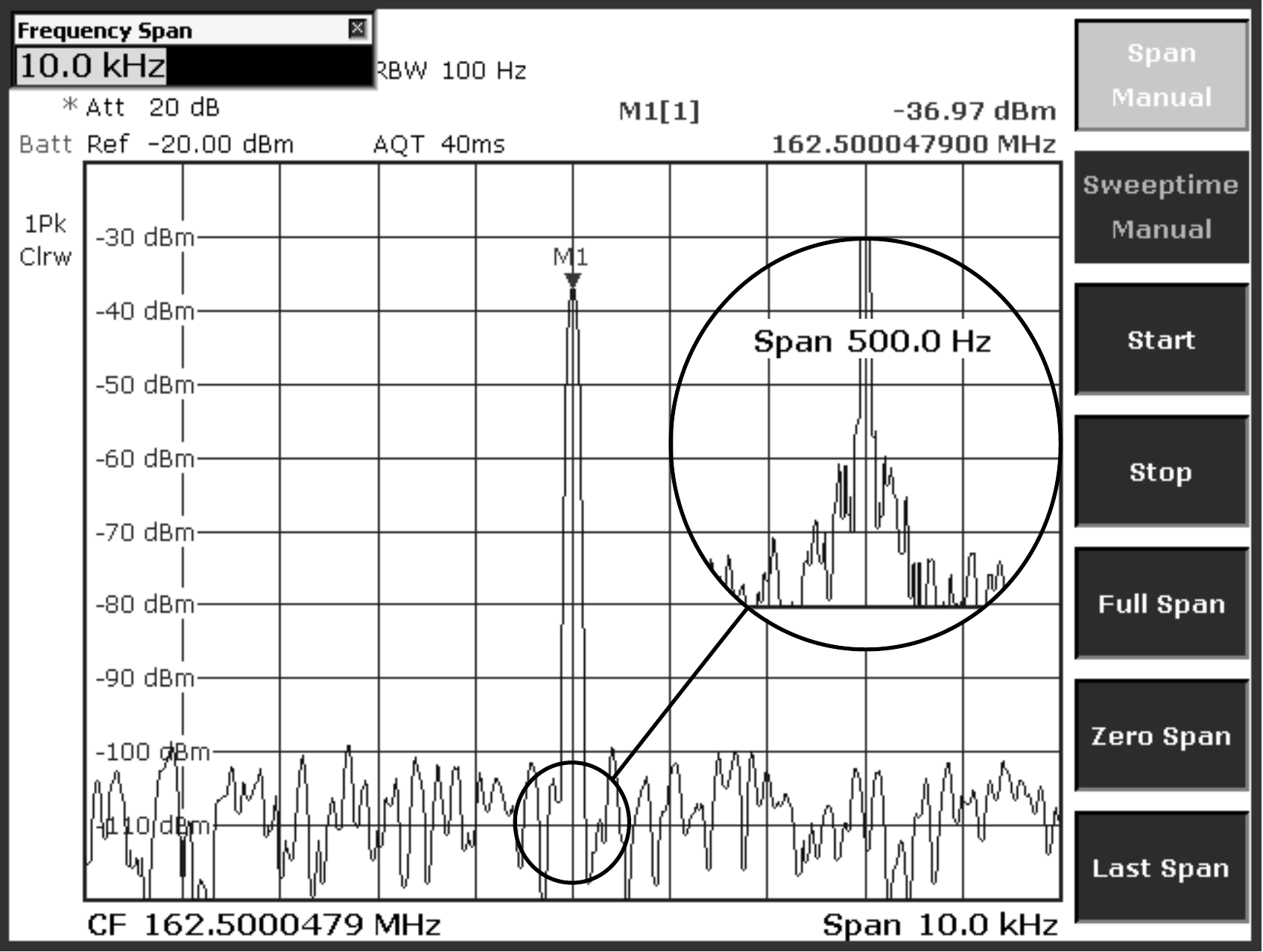}
\caption{Spectrum screenshot for buncher cavity monitoring signal while close-loop with front-end}
\label{fig:SpectrumScreenshot}       
\end{figure}

The feasibility and performance of the design were tested and evaluated by measuring basic parameters such as linearity, crosstalk, and phase noise for all expected frequencies, including $650 \,\mathrm{MHz}$, $325 \,\mathrm{MHz}$, and $162.5 \,\mathrm{MHz}$. Limited to the test conditions, the phase noise was measured on the LBNL cavity emulator only at 162.5 MHz, and the online closed-loop control performance was tested on the IMP 162.5 MHz buncher cavity.

\subsection{Standard parameters}

The instruments used in the standard parameter test and measurement in the laboratory were spectrum analyzers (SA, R\&S FSL), vector network analyzers (VNA, Agilent E5061A), and signal generators (SG, R\&S SMC100A).

\textbf{Linearity}: The measurement parameters of linearity are the coefficient of determination $R^2$ and P1dB. For down-conversion channels, the RF port input signal range is $-60 \sim 18 \,\mathrm{dBm}$, and for up-conversion channels, the IF port input signal range is $-60 \sim 10 \,\mathrm{dBm}$ to monitor the IF port for the down-conversion channel and RF port for the up-conversion channel amplitude with SA. As shown in Fig.~\ref{fig:Linearity} and Fig.~\ref{fig:P1dB}, the P1dB is higher than the design requirement, equivalent at $650 \,\mathrm{MHz}$ and $325 \,\mathrm{MHz}$ above $15 \,\mathrm{dBm}$, and slightly less optimal at $162.5 \,\mathrm{MHz}$. The coefficient of determination $R^2 > 99.991\%$ of the down-conversion channel at all expected frequency points. The linearity error was almost equivalent for all down-conversion channels at $650 \,\mathrm{MHz}$ from $1.00\% \sim -1.50\%$, at $325 \,\mathrm{MHz}$ from $0.75\% \sim -1.00\%$, and at $162.5 \,\mathrm{MHz}$ from $0.75\% \sim -1.50\%$.

\textbf{Crosstalk}: The measurement procedure employs a channel input signal at full scale to measure crosstalk at other channels, and $50 \,\mathrm{\Omega}$ matched at other unused ports, as results show in Fig.~\ref{fig:Crosstalk}. When working at the design's maximum operating point, the higher the frequency, the worse the crosstalk, $-60 \,\mathrm{dBm}$ at $650 \,\mathrm{MHz}$, $-68 \,\mathrm{dBm}$ at $325 \,\mathrm{MHz}$, and $-82 \,\mathrm{dBm}$ at $162.5 \,\mathrm{MHz}$. As described in~\cite{mavric2008design}, using the LO amplifier directly on the LO port increases the coupling between the channels and the near-center phase noise, thus, the leakage of the active mixer with the internal LO amplifier from the LO port to the RF port increases. Designing an additional amplifier and attenuator in series at the mixer LO port reduces leakage of RF signals to adjacent channels or from other channels through the LO distribution network. It is possible to generate the LOs and clocks from phase reference line or main oscillator~\cite{fu2010} for LLRF system on RF front-end which will include next version.

\textbf{S11 and SWR}: The reflection attenuation and the standing wave ratio (SWR) to evaluate the reflection were measured for the RF port. The S11 and SWR data were obtained by the VNA shown as in Fig.~\ref{fig:S11SWR1}. From the S11 test results, at $650\,\mathrm{MHz}$, $325\,\mathrm{MHz}$, and $162.5\,\mathrm{MHz}$, the reflected attenuation was better than $25\,\mathrm{dB}$, $35\,\mathrm{dB}$, and $45\,\mathrm{dB}$, respectively. The SWR test results indicate that the required frequency at the RF port is around 1.1:1, which is very close to the SWR of the cable; therefore, there is no excessive reflection in the cable introduced by the front-end affecting the measurement accuracy.

For the standard parameter measurement results, the linearity and P1dB are not the main problem for multi-frequency point supported design but rather crosstalk between adjacent channels is, although they have already carefully analyzed and designed in schematic and PCB layout. 

\subsection{Phase noise}

Phase noise and cavity emulator closed-loop experiments were performed at LBNL with 15/4 near IQ sampling at $162.5 \,\mathrm{MHz}$~\cite{doolittle2006digital}. The phase noise level test results for the loop of digital part, the loop of the digital and analog front-end, and the loop connected to the cavity emulator are shown in Fig.~\ref{fig:Phasenoise}. For the loop of the digital part, the cumulative phase noise is $0.006^\circ$ in the $10\,\mathrm{MHz}$ range, mainly contributed at higher frequencies, while the white noise is mainly below $1 \,\mathrm{MHz}$. For the digital and analog front-end loop, 1/f noise is the main noise at lower frequencies, the white noise inflection point appears at $1 \,\mathrm{kHz}$, and the cumulative phase noise is $0.002^\circ$/decades below $100 \,\mathrm{Hz}$. The growth becomes slow later, that is, the phase noise is mainly contributed at lower frequencies. For the loop after connecting the cavity emulator, the 1/f noise in the lower frequency part is significantly cancelled, with only some at the $100 \,\mathrm{Hz}$ position, as well as below $100 \,\mathrm{dBc}$. The cumulative phase noise is below $1\,\mathrm{MHz}$ and is less than $0.01^\circ$, and then becomes faster.

According to the test results, the main noise of the RF front-end within the designed control bandwidth is 1/f noise. The phase noise power spectrum is lower than $-80\,\mathrm{dBc}$ and the cumulative phase noise is $0.006^\circ$ in the $10\,\mathrm{MHz}$ range. The noise is mainly concentrated in the lower frequency range below $100\,\mathrm{Hz}$, and the higher frequency noise is mainly contributed by the digital part. In the connection cavity simulator (which can be regarded as a high Q bandpass filter), test results show that the cumulative phase noise lower than $0.01^\circ$ in the $10\,\mathrm{MHz}$ range and the phase noise power spectrum is lower than $-100\,\mathrm{dBc}$.

\subsection{Stability}

The online closed-loop experiment was performed on the CAFe $162.5 \,\mathrm{MHz}$ buncher cavity to evaluate the amplitude and the phase stability. The results are shown in Fig.~\ref{fig:B1Closeloop}. In the case of the beam load, the amplitude stability peak-to-peak value is $\pm 0.06\%$ and RMS value is $0.022\%$, the phase stability peak-to-peak value is $\pm 0.15^\circ$, and the corresponding RMS value is $0.044^\circ$, which can reach the design goal of $1\%$ and $0.1^\circ$ stability requirements for amplitude and phase, respectively. As can be seen clearly in Fig.~\ref{fig:B1Closeloop}, after then= moving average period of 10, the amplitude and phase short-term fluctuation was periodically controlled well by the LLRF controller, which was not caused by the beam operating in the pulse mode and at very low current during test processing.

The spectrum screenshot in Fig.~\ref{fig:SpectrumScreenshot} monitored by the phase monitor system (PMS) shows that there is still some noise on both sides (in the range around $\pm 50\,\mathrm{Hz}$) off the center frequency. After analyzing the LO generation mode between the RF front-end and PMS, it is clear that the noise may be caused by a $10\,\mathrm{MHz}$ source locked due to different LOs generated by two SGs, which is not introduced by the front-end. When the two sources are not locked, the phase rapidly changes within $\pm 180^\circ$ in a period around $50 \,\mathrm{Hz}$. Even in the case of the $10\,\mathrm{MHz}$ lock, the generated local oscillator signal will still cause long term slow fluctuation (hours), significantly affecting cavity phase stability. The $50 \,\mathrm{Hz}$ frequency of the fluctuation is likely introduced from the power frequency of the AC power supply.

\section{Conclusion}

The multi-frequency point supported front-end prototype is a modular, compact, and low maintenance solution for the CiADS linear accelerator LLRF RF front-end application, featuring a common structural design and compatibility with the MTCA.4 RTM standard. In the design, the linearity of each frequency and the crosstalk suppression between channels are considered carefully. The test results show that the broadband RF front-end design can meet the requirements of CiADS for high stability of phase and amplitude at different frequencies. Considering the other requirements of CiADS for other indicators such as reliability, the main improvement should be designed to further reduce the signal leakage between channels and test all frequencies online or at least on cavity emulators. Another improvement should be the up and down IF frequency separation, which can further reduce the IF crosstalk performance between the up-conversion and down-conversion channels; that and other components or designs with limited aspects can also be improved in the next version. Concerning availability, maintainability, high performance, and modular and standardized design, the batched deliverable of a general-purpose multi-frequency point supported RF front-end for the LLRF system that can reduce the complicated manufacturing and maintenance for multi-frequency point accelerator application is achieved.

\begin{acknowledgements}
The multi-frequency supported front-end prototype was guided and supported by Lawrence Doolittle, Gang Huang, Derun Li, and John Byrd from LBNL, and benefited from Lawrence Doolittle’s and Gang Huang's valuable comments and suggestions on the design. Thanks to Derun Li and John Byrd's project financial support.
\end{acknowledgements}

\end{document}